\documentclass[aps,prl,twocolumn,superscriptaddress]{revtex4}
\usepackage{graphicx}
\usepackage{dcolumn}
\usepackage{bm}
\usepackage{color}
\usepackage{hyperref}
\usepackage{amsmath}
\usepackage{amssymb}

\begin{document}

\title{Continuous growth of droplet size variance due to condensation in turbulent clouds}

\author{Gaetano Sardina}\email[]{gaetano.sardina@misu.su.se}
\affiliation{Department of Meteorology and SeRC, Stockholm University, Stockholm, Sweden}
\author{Francesco Picano}
\affiliation{Linn{\'e} FLOW Centre and SeRC, KTH Mechanics, Stockholm, Sweden}
\affiliation{Department of Industrial Engineering, University of Padova, Padova, Italy}
\author{Luca Brandt}
\affiliation{Linn{\'e} FLOW Centre and SeRC, KTH Mechanics, Stockholm, Sweden}
\author{Rodrigo Caballero}
\affiliation{Department of Meteorology and SeRC, Stockholm University, Stockholm, Sweden}

\date{\today}

\begin{abstract}

  We use a stochastic model and direct numerical simulation to study the impact of
  turbulence on cloud droplet growth by condensation. We show that the variance of the
  droplet size distribution increases in time as $t^{1/2}$, with growth rate
  proportional  \textcolor{black}{to} the large-to-small turbulent scale separation and to the turbulence
  integral scales but independent of the \textcolor{black}{mean turbulent dissipation}. Direct numerical
  simulations confirm this result and produce realistically broad droplet size spectra
  over time intervals of 20 minutes, comparable with the time of rain formation.
\end{abstract}


\maketitle

Many multi-scale processes---including nutrient foraging of plankton, gas/dust accretion
disks in astrophysics and spray evaporation and combustion in engines
\cite{stock,planet,comb,bec1}---involve the interaction between small particles and
tracers transported in a turbulent flow. Here we focus on the case of droplet
condensation in turbulent warm (i.e.\ ice-free) clouds.  Clouds are a leading source
of uncertainty in climate modeling \cite{shio} due to the difficulty of accurately
parameterizing the macro-scale effects of micro-scale physical processes, such as the
effect of droplet size distribution on precipitation rates and cloud albedos.  

The role of turbulence in cloud microphysics presents a range of open questions
\cite{shaw,bode,grar}, particularly as a possible solution for the
  ``bottleneck'' problem of rain formation. All cloud droplet populations evolve through
  a sequence of steps: (1) nucleation/activation of cloud condensation nuclei, (2)
  droplet growth by condensation and (3) growth to raindrop size by collision and
  coalescence.  The passage from (2) to (3) presents a bottleneck because collisional
  growth is only triggered when the droplet population acquires a sufficiently broad
  size distribution, but condensational growth is inversely proportional to droplet
  radius 
   \textcolor{black}{which} intrinsically tends to narrow the size distribution. Nonetheless,
  warm clouds are routinely observed to precipitate within $\sim$20 minutes of
  formation. Understanding the mechanisms that break the condensational bottleneck is a
  longstanding and still unresolved problem in atmospheric physics.
  
  Turbulence has often been invoked as a key process in this context since it can
  enhance collision rates via inertial clustering \cite{collins,bec2} and the so called
  ``sling effect'' \cite{falko}. Turbulence also induces fluctuations in the
  supersaturation field that can potentially broaden droplet spectra in the
  condensational stage \cite{grar}. Early studies using analytical models
  \cite{coop,kulmala} and direct numerical simulations (DNS) \cite{vail} generally
  showed too little broadening as compared with observations \cite{breng}. Later work
  attempting to simulate large-scale turbulence in an $O$(100 m) cloud \cite{celani,mazzino,paoli}
  showed a dramatic broadening of the droplet size spectrum but only with ad-hoc
  assumptions about the small-scale supersaturation fluctuations.  Lanotte at al.\
  \cite{lanotte} modelled both small- and large-scale effects on the droplet size
  distribution with simulations of a cloud of 70 cm and pointed out the importance the
  Taylor Reynolds number, $Re_\lambda$, the non dimensional parameter measuring the
  large to small scale separation in homogeneous isotropic turbulence. In particular,
  they suggested that droplet spectral broadening should scale with $Re_\lambda$.

  A question that has not been addressed so far is the long-term fate of the droplet
  spectrum: does it reach a steady state, or does it continue to evolve? The large range
  of scales involved makes DNS very computationally demanding, and all existing
  simulations consider time spans between few seconds and two minutes, well below the
  observed rain formation timescale.

  Here, we approach this question by first deriving an analytical expression for the
  standard deviation of the droplet radius squared (droplet surface area) in terms of
  the thermodynamics and turbulence characteristics, modeling the fluctuations as
  stochastic processes. We demonstrate that the droplet size distribution 
  increases monotonically with time as $t^{1/2}$. We validate this analytical result by
  extending previous numerical results with DNS and large eddy simulations (LES) to
  timescales comparable with those of rain formation, about 20 minutes. Our results
  imply that every warm cloud would precipitate if given enough time.  The broadening
  rate is determined by the large scale turbulent motions and by the positive
  correlation between droplet surface area and local supersaturation.

 
Our physical model is 
analogous to that in \cite{lanotte}, \textcolor{black}{a detailed description can be found in Supplemental Materials \cite{suppl}.} 
\textcolor{black}{We assume homogeneous isotropic turbulence obeying the incompressible Navier-Stokes equations}
\begin{align}
\frac{\partial {\bf u}}{\partial t}+{\bf u}\cdot\nabla {\bf u}&=-\frac{\nabla p}{\rho}+\nu \nabla^2 {\bf u}+ {\bf f}, \label{fluid_u}
\end{align}
where ${\bf u}$ is the divergence-free fluid velocity, $p$ the pressure, $\rho$ the air
density, ${\bf f}$ an external forcing to maintain a statistically stationary state and
$\nu$ the kinematic viscosity. \textcolor{black}{This approximation is valid for clouds smaller than $L
\approx 100$~m, which allow  spatial inhomogeneity and large scale variations of
the thermodynamic parameters to be safely neglected}.  The supersaturation field $s$ is transported by the fluid
according to
\begin{align}
\frac{\partial s}{\partial t}+{\bf u}\cdot\nabla s &= \kappa \nabla^2 s + A_1 w -\frac{s}{\tau_s}, \label{fluid_s}
\end{align}
a generalization of the Twomey model \cite{twom}. The diffusivity of the water vapor in
air is denoted by $\kappa$, $w$ is the velocity component in the gravity direction, $A_1
w$ is a source/sink term of supersaturation resulting from the variation in temperature
and pressure with height. The supersaturation relaxation time $\tau_s$ depends on
droplet concentration and dimensions \cite{prup}, $\tau_s^{-1}=4\pi \rho_w A_2 A_3 \sum
R_i/V$ where $R_i$ are the radii of the droplets in the volume $V$, $\rho_w$ the water
density, $A_1$, $A_2$ and $A_3$ thermodynamic coefficients
\cite{lanotte}. 
The droplets are assumed to behave as 
rigid spheres smaller than the Kolmogorov turbulent scale, at low mass fraction to
neglect feedback on the flow. The only forces governing the droplet motion are gravity
and the Stokes drag (nucleation/activation is not considered):
\begin{align}
\frac{d {\bf v_d}}{dt}&=\frac{{\bf u}({\bf x_d},t)-{{\bf v_d}}}{\tau_d}-g {\bf e_z}, \quad 
\frac{d {\bf x_d}}{dt}={\bf v_d}  \label{space_d}
\end{align}
with ${\bf x_d}$ and ${\bf v_d}$ the droplet position and velocity, ${\bf u}({\bf
  x_d},t)$ the fluid velocity at droplet position, $\tau_d=2\rho_w R_i^2/(9 \rho\nu)$
the droplet relaxation time, $g$ the gravitational acceleration.  The supersaturation at
the droplet position, $s({\bf x_d},t)$, determines the droplet evolution via
\begin{align}
\frac{d { R_i^2}}{dt}&=2 A_3 {s({\bf x_d},t)} \label{rad_d}
\end{align} 
An exact equation for the average of the droplet radius fluctuations is obtained from (\ref{rad_d}), 
\begin{equation}
\frac{d { \langle (R_i^{2'})^2\rangle}}{dt}=\frac{d \sigma_{R^2}^2}{dt}=4 A_3 \langle s' R^{2'}\rangle \label{exact}
\end{equation}
showing that $\langle (R_i^{2'})^2\rangle$ increases linearly with time only if the correlation $\langle s' R^{2'}\rangle$ reaches a non-zero statistical steady state.

To quantitatively estimate the droplet growth, we adopt a 1-D stochastic model for the
velocity fluctuations $w_i$ and the supersaturation field $s'_i$ of the i-th droplet.
Such an approach implicitly assumes that the small-scale turbulent motions have a
negligible influence on the macroscopic observables. This assumption is motivated by
previous results revealing the broadening of the droplet size distribution with
$Re_\lambda$ \cite{lanotte} and fully justified a posteriori by the numerical \textcolor{black}{simulations}
presented below.

Homogeneous isotropic turbulence and supersaturation are modelled with two Langevin equations \cite{pope}:
\begin{equation}
 w'_i(t+dt)=w'_i(t)-\frac{w'_i(t)}{T_0}dt+v_{rms}\sqrt{2\frac{dt}{T_0}}\xi_{i}(t),\label{1d_u} \\
\end{equation}
where \textcolor{black}{$v_{rms}$ is the root mean square of the turbulent velocity fluctuations}, $\xi(t)$ is \textcolor{black}{a zero-mean Gaussian white noise, nondimensionalized in order 
to be $\delta$-correlated in time} and $T_0$ the turbulence integral time scale\textcolor{black}{;} and 
\begin{align}
 s'_i(t+dt)=s'_i(t)-\frac{s'_i}{T_0}dt+A_1w'_idt-\frac{s'_i}{\langle \tau_{s}\rangle}dt+\nonumber \\
 \sqrt{(1-C_{ws}^2) \langle s'^2 \rangle \frac{2dt}{T_0}}\eta_i(t)+C_{ws} \sqrt{\langle s'^2\rangle \frac{2dt}{T_0}}\xi_{i}(t)      \label{1d_s}
\end{align}
for the supersaturation with forcing from the velocity field.
Here $C_{ws}=\langle w's'\rangle/(v_{rms}\sqrt{\langle s'^2\rangle})$ is the normalized velocity-supersaturation correlation,
 $\langle \tau_s \rangle$ is the supersaturation relaxation time based on the mean droplet radius and
$\eta(t)$ \textcolor{black}{zero-mean Gaussian white noise}, $\delta$-correlated
in time. Equation (\ref{1d_s}) represents a stochastic version of
\textcolor{black}{the} Twomey model.
The mean updraft is zero as the mean supersaturation (the mean droplet radius does not change); 
entrainment effects \cite{schum}, collisions and inhomogeneities are also not considered to analyze the conservative case when the droplet spectral broadening is only induced by supersaturation fluctuations. 

From (\ref{rad_d}), (\ref{1d_u}) and (\ref{1d_s}), assuming $\langle \tau_s \rangle \ll T_0$ as in real clouds, the fluctuation correlations become 
\begin{align}
 \frac{d\langle s' R^{2'}\rangle}{dt}=A_1\langle w'R^{2'}\rangle+2A_3\langle s'^2 \rangle-\frac{\langle s'R^{2'}\rangle}{\langle \tau_{s}\rangle} \label{1d_sr2}\\
 \frac{d\langle w' R^{2'}\rangle}{dt}=2A_3\langle w's' \rangle-\frac{\langle w'R^{2'}\rangle}{T_0}\label{1d_wr2}\\
  \frac{d\langle s^{'2}\rangle}{dt}=2A_1\langle w's' \rangle-2\frac{\langle s'^2\rangle}{\langle \tau_{s}\rangle}\label{1d_s2}\\
    \frac{d\langle w's'\rangle}{dt}=A_1 v_{rms}^2-\frac{\langle w' s'\rangle }{\langle \tau_{s}\rangle}.\label{1d_ws}
\end{align}
Assuming \textcolor{black}{statistically quasi-steady state conditions} ($\frac{d\langle
    \rangle}{dt}=0$) we find that $ \langle s'^2 \rangle_{qs}= A_1^2 v_{rms}^2 {\langle
    \tau_{s}\rangle^2}\label{s2_qs}$ and
\begin{align}
  \langle s' R^{2'}\rangle_{qs}=2A_3A_1^2 v_{rms}^2 {\langle \tau_{s}\rangle^2}T_0= 2A_3 \langle s'^2 \rangle_{qs}T_0\label{sr2_qs}
\end{align}
which give, using (\ref{exact}), 
\begin{align}
  \sigma_{R^2}= \sqrt{8} A_3 A_1  v_{rms} {\langle \tau_{s}\rangle} (T_0 t)^{1/2}=\sqrt{8\langle s'^2 \rangle_{qs}} A_3 (T_0 t)^{1/2}  \label{sigr2}.
\end{align}
The model shows that  the 
variance of the droplet distribution increases monotonically in a turbulent flow
 even \textcolor{black}{though}  the supersaturation fluctuations  \textcolor{black}{reach} a 
statistical steady state  $s_{qs}$. 
The correlation $\langle s' R^{2'}\rangle_{qs}$,  \textcolor{black}{which is} directly responsible for the radius growth rate, is proportional to the level of fluctuations of the supersaturation field and to the integral scale of the turbulence, see (\ref{sr2_qs}).
Expression (\ref{sigr2}) can be formulated in terms of Kolmogorov scales since $v_{rms}\simeq Re_\lambda^{1/2}v_\eta$ and $T_0\simeq 0.06 Re_\lambda \tau_\eta$ \cite{pope}:
\begin{equation}
\sigma_{R^2}\simeq 0.7 A_3 A_1  \nu^{1/2} {\langle \tau_{s}\rangle} Re_\lambda t^{1/2}  \label{sigr2_2}.
\end{equation}
Note that for $t=T_0$ (short  \textcolor{black}{compared with rain formation time}) the lower limit proposed in \cite{lanotte} is
recovered, $\sigma_{R^2}\simeq Re_\lambda^{3/2}$.  From (\ref{sigr2_2}) we note that
$\sigma_{R^2}$ at a fixed time depends only on the scale separation represented by
$Re_\lambda$ and not on the value of the mean dissipation inside the clouds. This
implies that clouds with different dissipation rate and same Reynolds number have an
identical behavior in terms of droplet growth by condensation. The droplet/turbulence
condensation dynamics does not depend on the turbulent small scales: the correlation
between the supersaturation field and the droplet surface area, governing the
distribution broadening, is determined by the large flow scales.  This result is in
contrast with the belief that the variance of the droplet distribution should not grow
indefinitely as turbulence tends to decorrelate the particle size from the local
saturation field \cite{grar}.

\begin{table}[b]
\begin{center}
\begin{tabular}{ c  c  c  c  c  c  c c }
\hline
\hline
Label & $N^3$ & $L_{box}$ & $v_{rms}$ & $T_{L}$ & $T_{0}$ &  
$Re_\lambda$ & $N_{d}$ \\
 &   &$[m]$ & $[m/s]$ & $[s]$ & $[s]$ &  
$ $ & $ $ \\
\hline
DNS A1/2 & $64^3$ & $0.08$ & $0.035$ & $2.3$ & $0.64$&$45$ & $6\times 10^4$\\
DNS B1/2 & $128^3$ & $0.2$ & $0.05$ & $4$ & $0.95$ & $95$ & $9.8\times 10^5$\\
DNS C1/2 & $256^3$ & $0.4$ & $0.066$ & $6$ & $1.5$ & $150$ & $9\times 10^6$\\
DNS D1 & $1024^3$ & $1.5$ & $0.11$ & $14$ & $3$  & $390$ & $4.4\times 10^8$\\
DNS E1 & $2048^3$ & $3$ & $0.12$ & $30$ & $4$  & $600$ & $3.\times 10^9$\\
LES F1 & $512^3$ & $100$ & $0.7$ & $142$ & $33$    &$5000$ & $1.3\times 10^{14}$\\
\hline
\end{tabular}
\caption{Parameters of the simulations. The resolution $N$,  the cloud size $L_{box}$,  the root mean square of the turbulent velocity fluctuations $v_{rms}$, and $T_L=L_{box}/v_{rms}$ an approximation of the large turbulent scales. $T_0$ indicates  the
integral time $T_0=(\pi/2v_{rms}^3)\int  [E(k)/k]dk$ with $k$ the wavenumber and $E(k)$ the turbulent kinetic energy spectra \cite{pope}.  The total number of droplets is indicated by $N_d$.}
\label{tab_1}
\end{center}
\end{table}

To test our predictions, we run simulations by gradually increasing the size of the
computational clouds from few centimeters to 100~m.  The governing equations (1-4)
are solved with a classical pseudo-spectral code for the fluid phase coupled with a
Lagrangian algorithm for the droplets \cite{cj}.  All cases share the same turbulent
kinetic energy dissipation $\varepsilon=10^{-3}$ m$^2$s$^{-3}$, a value typically measured in
stratocumuli. This corresponds to the same small-scale dynamics, with Kolmogorov scale
$\eta=(\nu^3/\varepsilon)^{1/4}\approx1$ mm, Kolmogorov time
$\tau_\eta=(\nu/\varepsilon)^{1/2} \approx 0.1$ s and velocity
$v_\eta=\eta/\tau_\eta\approx 1$ cm/s. We examine droplets with 2 different initial
radii, $13 \;\mu m$ and $5\; \mu m$, denoted as case 1 and 2, with supersaturation
relaxation time $\tau_s=2.5$ and 7 s, and same concentration (130 droplets per cm$^3$).
The reference temperature and pressure are $\theta_0=283$ K and $P=10^5$ Pa, with 
$A_1=5\times10^{-4}$ m$^{-1}$, $A_2=350$ m$^3$/kg, $A_3=50\;\mu$m$^2$/s. 
The simulation parameters are reported in table \ref{tab_1}. 
Note that simulation DNS E1, \textcolor{black}{carrying order $10^9$ droplets, is to the
  best of our knowledge the largest Direct Numerical Simulation of a warm cloud ever
  performed.}

\begin{figure}[t]
\centering
\includegraphics[width =.49 \textwidth]{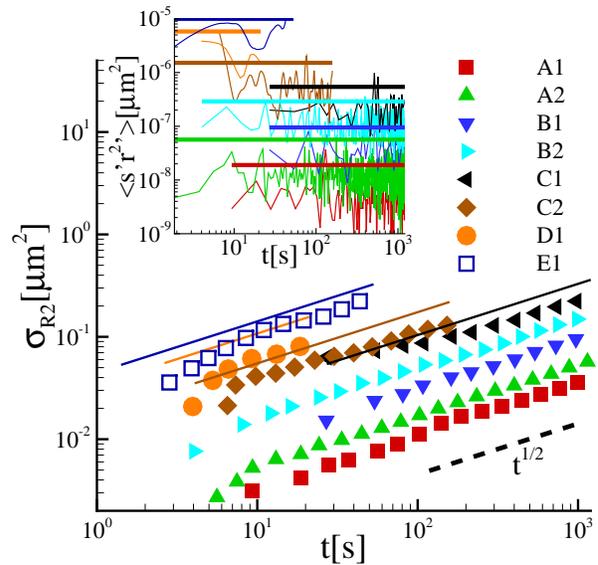}
\caption{Root mean square of the square droplet radius fluctuations $\sigma_{R^2}$
  versus time from simulations (symbols) and the prediction of the stochastic model
  (\ref{sigr2}) (lines). Inset: correlation $\langle s' R^{2'}\rangle$ from simulations
  (thin lines) and from Equation (\ref{sr2_qs}) (thick lines).}
\label{fig_1}
\end{figure}

The time evolution of $\sigma_{R^2}=\sqrt{\langle (R^{2'})^2\rangle}$ is shown in Figure
\ref{fig_1} for all cases investigated.  The data confirm the predictions from
(\ref{sigr2}), i.e.\ that $\sigma_{R^2} \propto t^{1/2}$.

The correlation $\langle s' R^{2'}\rangle$ is displayed in the inset of Figure
\ref{fig_1} (thin solid line): In all cases, $\langle s' R^{2'}\rangle$
\textcolor{black}{reaches} a statistical steady state, fluctuating around the value
determining the growth of $\sigma_{R^2}$. The turbulence creates a positive correlation
between supersaturation and droplet surface area fluctuations that increases  \textcolor{black}{as
the turbulent scale separation---i.e.\ the cloud size---increases}. The agreement between
the model and the numerical data is remarkable for the largest domain sizes where scale
separation is significant and viscous effects can be neglected. For small
$Re_{\lambda}$, viscous effects are important and the stochastic inviscid model
overestimates the correct behavior. \textcolor{black}{For a detailed comparison between
  DNS and stochastic model see Supplemental Materials \cite{suppl}}.

\begin{figure}[t]
\centering
\includegraphics[width =.49 \textwidth]{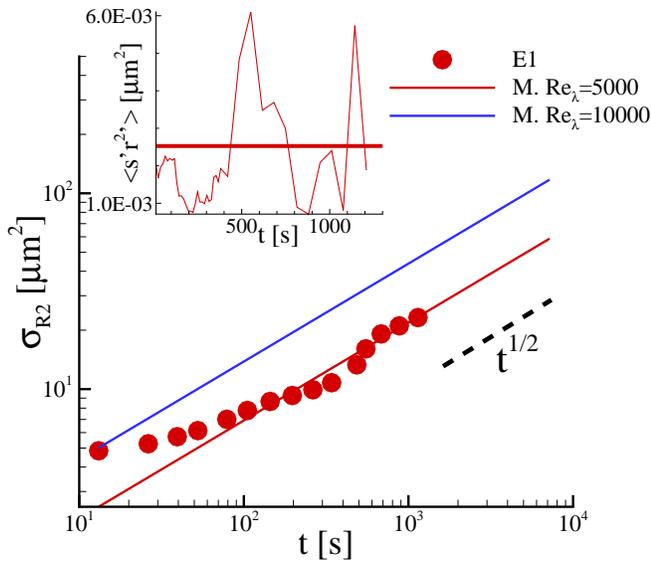}
\caption{Root main square of the square droplet radius fluctuations $\sigma_{R^2}$ versus time for the LES simulations (symbols) and the model (\ref{sigr2}) (lines). Inset: correlation $\langle s' R^{2'}\rangle$  from the LES (thin line) and the model (\ref{sr2_qs}) (denoted by "M." in the legend-thick lines).}
\label{fig_3}
\end{figure}

To test the model for a larger cloud size, we perform a large eddy simulation (LES F1)
of a cloud of about 100~m size. LES can be seen as a good model for our problem since it
fully resolves the larger flow scale, those relevant to droplet
condensation/evaporation, as shown above. \textcolor{black}{For numerical details see Supplemental Materials \cite{suppl}}. The Taylor Reynolds number is 5000.
The time evolution of $\sigma_{R^2}$ and of $\langle s' R^{2'}\rangle$ are depicted in Fig.
\ref{fig_3} together with the analytical predictions from (\ref{sr2_qs}) and
(\ref{sigr2}). The agreement between the two fully validates our model.

\begin{figure}[b!]
\centering
\includegraphics[width =.49 \textwidth]{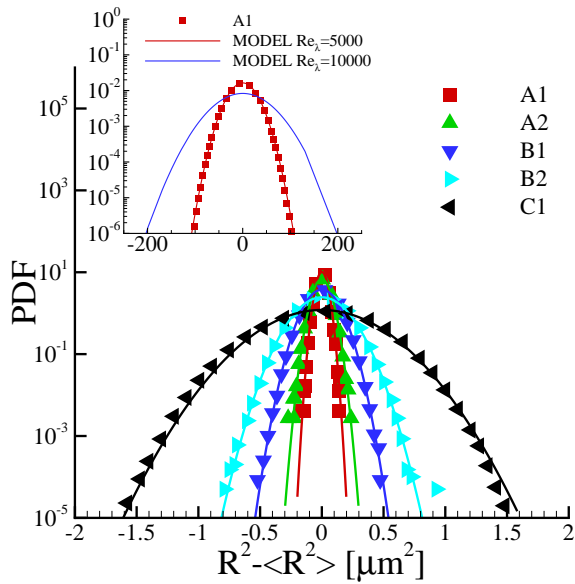}
\caption{Probability density functions (PDFs) of the square radius fluctuations after a simulation time corresponding to about 20 minutes (symbols). The lines represent Gaussian distributions with the same variance. The inset reports data from the LES of a large cloud.}
\label{fig_2}
\end{figure}

To motivate the use of the variance $\sigma_{R^2}^2$ to define the droplet size
distribution we show in Figure \ref{fig_2} that its probability distribution
is Gaussian, in agreement with measurements in real stratocumuli \cite{ditas}.  The data
in the figure refer to the final simulation time (about 20
minutes) and are compared to Gaussian distributions of equal variance\textcolor{black}{; note that error bars are about the same size as the plotting symbols and
  not visible in the plot}.
The size distribution from the LES of the large cloud (see inset) reveals that the
Gaussian can be fitted just using the value of $\sigma_{R^2}$ from the  \textcolor{black}{stochastic} model
also at this higher Reynolds number.

 

In summary, we have derived an analytical expression for the role of turbulence on the
dynamics of droplet condensation and validated it against large-scale numerical
simulations.  We show that the standard deviation of the square droplet radius
fluctuations, $\sigma_{R^2}$, increases in time as $t^{1/2}$; the growth rate depends
linearly on the turbulent scale separation parametrized by $Re_\lambda$.  As shown in
Figure~2, for a cloud with $Re_\lambda=10000$---a typical value estimated in cumuli with
integral scale of 100~m \cite{grar}---our expression predicts that $\sigma_{R^2}$
reaches values in line with observations in real clouds (see \cite{breng}, their
Figure~4) on timescales of less than 20 minutes.

The stochastic approach proposed here may be  \textcolor{black}{generalized} to consider additional physics and
adapted to different micro-scale phenomena in turbulent environments; this may also
require a numerical solution of the governing Langevin equations, something still order
of magnitudes cheaper than a fully-resolved DNS.  Indeed our analytical relation
predicts numerical results requiring $10^{17}$ degrees of freedom. From a practical
viewpoint, this indicates the promising potential of modeling approaches based on PDFs.

Our results  \textcolor{black}{represent} a lower limit for the impact of turbulence on warm rain formation
since real clouds typically exceed 100~m in scale and are in general non-homogeneous, \textcolor{black}{featuring a fluctuating temperature field
and vigorous entrainment of relatively dry air from outside the cloud leading to
enhanced supersaturation fluctuations within the cloud.}  These additional effects would
lead to even larger values of $\sigma_{R^2}$, more than sufficient to explain the
spectral broadening observed in real clouds.


\begin{thebibliography}{32}
\expandafter\ifx\csname natexlab\endcsname\relax\def\natexlab#1{#1}\fi
\expandafter\ifx\csname bibnamefont\endcsname\relax
  \def\bibnamefont#1{#1}\fi
\expandafter\ifx\csname bibfnamefont\endcsname\relax
  \def\bibfnamefont#1{#1}\fi
\expandafter\ifx\csname citenamefont\endcsname\relax
  \def\citenamefont#1{#1}\fi
\expandafter\ifx\csname url\endcsname\relax
  \def\url#1{\texttt{#1}}\fi
\expandafter\ifx\csname urlprefix\endcsname\relax\def\urlprefix{URL }\fi
\providecommand{\bibinfo}[2]{#2}
\providecommand{\eprint}[2][]{\url{#2}}

\bibitem[{\citenamefont{Taylor and Stocker}(2012)}]{stock}
\bibinfo{author}{\bibfnamefont{J.~R.} \bibnamefont{Taylor}} \bibnamefont{and}
  \bibinfo{author}{\bibfnamefont{R.}~\bibnamefont{Stocker}},
  \bibinfo{journal}{Science} \textbf{\bibinfo{volume}{338}},
  \bibinfo{pages}{675} (\bibinfo{year}{2012}).

\bibitem[{\citenamefont{Johansen et~al.}(2007)\citenamefont{Johansen, Oishi,
  Mac~Low, Klahr, Henning, and Youdin}}]{planet}
\bibinfo{author}{\bibfnamefont{A.}~\bibnamefont{Johansen}},
  \bibinfo{author}{\bibfnamefont{J.~S.} \bibnamefont{Oishi}},
  \bibinfo{author}{\bibfnamefont{M.-M.} \bibnamefont{Mac~Low}},
  \bibinfo{author}{\bibfnamefont{H.}~\bibnamefont{Klahr}},
  \bibinfo{author}{\bibfnamefont{T.}~\bibnamefont{Henning}}, \bibnamefont{and}
  \bibinfo{author}{\bibfnamefont{A.}~\bibnamefont{Youdin}},
  \bibinfo{journal}{Nature} \textbf{\bibinfo{volume}{448}},
  \bibinfo{pages}{1022} (\bibinfo{year}{2007}).

\bibitem[{\citenamefont{Jenny et~al.}(2012)\citenamefont{Jenny, Roekaerts, and
  Beishuizen}}]{comb}
\bibinfo{author}{\bibfnamefont{P.}~\bibnamefont{Jenny}},
  \bibinfo{author}{\bibfnamefont{D.}~\bibnamefont{Roekaerts}},
  \bibnamefont{and}
  \bibinfo{author}{\bibfnamefont{N.}~\bibnamefont{Beishuizen}},
  \bibinfo{journal}{Prog. Energ. Combust.} \textbf{\bibinfo{volume}{38}},
  \bibinfo{pages}{846} (\bibinfo{year}{2012}).

\bibitem[{\citenamefont{Bec et~al.}(2014{\natexlab{a}})\citenamefont{Bec,
  Homann, and Krstulovic}}]{bec1}
\bibinfo{author}{\bibfnamefont{J.}~\bibnamefont{Bec}},
  \bibinfo{author}{\bibfnamefont{H.}~\bibnamefont{Homann}}, \bibnamefont{and}
  \bibinfo{author}{\bibfnamefont{G.}~\bibnamefont{Krstulovic}},
  \bibinfo{journal}{Phys. Rev. Lett.} \textbf{\bibinfo{volume}{112}},
  \bibinfo{pages}{234503} (\bibinfo{year}{2014}{\natexlab{a}}).

\bibitem[{\citenamefont{Shiogama and Ogura}(2014)}]{shio}
\bibinfo{author}{\bibfnamefont{H.}~\bibnamefont{Shiogama}} \bibnamefont{and}
  \bibinfo{author}{\bibfnamefont{T.}~\bibnamefont{Ogura}},
  \bibinfo{journal}{Nature} \textbf{\bibinfo{volume}{505-7481}},
  \bibinfo{pages}{34} (\bibinfo{year}{2014}).

\bibitem[{\citenamefont{Shaw}(2003)}]{shaw}
\bibinfo{author}{\bibfnamefont{R.~A.} \bibnamefont{Shaw}},
  \bibinfo{journal}{Ann. Rev. Fluid Mech.} \textbf{\bibinfo{volume}{35}}
  (\bibinfo{year}{2003}).

\bibitem[{\citenamefont{Bodenschatz et~al.}(2010)\citenamefont{Bodenschatz,
  Malinowski, Shaw, and Stratmann}}]{bode}
\bibinfo{author}{\bibfnamefont{E.}~\bibnamefont{Bodenschatz}},
  \bibinfo{author}{\bibfnamefont{S.~P.} \bibnamefont{Malinowski}},
  \bibinfo{author}{\bibfnamefont{R.~A.} \bibnamefont{Shaw}}, \bibnamefont{and}
  \bibinfo{author}{\bibfnamefont{F.}~\bibnamefont{Stratmann}},
  \bibinfo{journal}{Science} \textbf{\bibinfo{volume}{327}},
  \bibinfo{pages}{970} (\bibinfo{year}{2010}).

\bibitem[{\citenamefont{Grabowski and Wang}(2013)}]{grar}
\bibinfo{author}{\bibfnamefont{W.~W.} \bibnamefont{Grabowski}}
  \bibnamefont{and} \bibinfo{author}{\bibfnamefont{L.-P.} \bibnamefont{Wang}},
  \bibinfo{journal}{Ann. Rev. Fluid Mech.} \textbf{\bibinfo{volume}{45}},
  \bibinfo{pages}{293} (\bibinfo{year}{2013}).

\bibitem[{\citenamefont{Sundaram and Collins}(1997)}]{collins}
\bibinfo{author}{\bibfnamefont{S.}~\bibnamefont{Sundaram}} \bibnamefont{and}
  \bibinfo{author}{\bibfnamefont{L.~R.} \bibnamefont{Collins}},
  \bibinfo{journal}{J. Fluid Mech.} \textbf{\bibinfo{volume}{335}},
  \bibinfo{pages}{75} (\bibinfo{year}{1997}).

\bibitem[{\citenamefont{Bec et~al.}(2014{\natexlab{b}})\citenamefont{Bec,
  Homann, and Ray}}]{bec2}
\bibinfo{author}{\bibfnamefont{J.}~\bibnamefont{Bec}},
  \bibinfo{author}{\bibfnamefont{H.}~\bibnamefont{Homann}}, \bibnamefont{and}
  \bibinfo{author}{\bibfnamefont{S.~S.} \bibnamefont{Ray}},
  \bibinfo{journal}{Phys. Rev. Lett.} \textbf{\bibinfo{volume}{112}},
  \bibinfo{pages}{184501} (\bibinfo{year}{2014}{\natexlab{b}}).

\bibitem[{\citenamefont{Falkovich et~al.}(2002)\citenamefont{Falkovich, Fouxon,
  and Stepanov}}]{falko}
\bibinfo{author}{\bibfnamefont{G.}~\bibnamefont{Falkovich}},
  \bibinfo{author}{\bibfnamefont{A.}~\bibnamefont{Fouxon}}, \bibnamefont{and}
  \bibinfo{author}{\bibfnamefont{M.}~\bibnamefont{Stepanov}},
  \bibinfo{journal}{Nature} \textbf{\bibinfo{volume}{419}},
  \bibinfo{pages}{151} (\bibinfo{year}{2002}).

\bibitem[{\citenamefont{Cooper}(1989)}]{coop}
\bibinfo{author}{\bibfnamefont{W.~A.} \bibnamefont{Cooper}},
  \bibinfo{journal}{J. Atmos. Sci.} \textbf{\bibinfo{volume}{46}},
  \bibinfo{pages}{1301} (\bibinfo{year}{1989}).

\bibitem[{\citenamefont{Kulmala et~al.}(1997)\citenamefont{Kulmala, Rannik,
  Zapadinsky, and Clement}}]{kulmala}
\bibinfo{author}{\bibfnamefont{M.}~\bibnamefont{Kulmala}},
  \bibinfo{author}{\bibfnamefont{{\"U}.}~\bibnamefont{Rannik}},
  \bibinfo{author}{\bibfnamefont{E.~L.} \bibnamefont{Zapadinsky}},
  \bibnamefont{and} \bibinfo{author}{\bibfnamefont{C.~F.}
  \bibnamefont{Clement}}, \bibinfo{journal}{Journal of Aerosol Science}
  \textbf{\bibinfo{volume}{28}}, \bibinfo{pages}{1395} (\bibinfo{year}{1997}).

\bibitem[{\citenamefont{Vaillancourt et~al.}(2001)\citenamefont{Vaillancourt,
  Yau, and Grabowski}}]{vail}
\bibinfo{author}{\bibfnamefont{P.}~\bibnamefont{Vaillancourt}},
  \bibinfo{author}{\bibfnamefont{M.}~\bibnamefont{Yau}}, \bibnamefont{and}
  \bibinfo{author}{\bibfnamefont{W.}~\bibnamefont{Grabowski}},
  \bibinfo{journal}{J. Atmos. Sci.} \textbf{\bibinfo{volume}{58}},
  \bibinfo{pages}{1945} (\bibinfo{year}{2001}).

\bibitem[{\citenamefont{Brenguier and Chaumat}(2001)}]{breng}
\bibinfo{author}{\bibfnamefont{J.-L.} \bibnamefont{Brenguier}}
  \bibnamefont{and} \bibinfo{author}{\bibfnamefont{L.}~\bibnamefont{Chaumat}},
  \bibinfo{journal}{J. Atmos. Sci.} \textbf{\bibinfo{volume}{58}},
  \bibinfo{pages}{628} (\bibinfo{year}{2001}).

\bibitem[{\citenamefont{Celani et~al.}(2007)\citenamefont{Celani, Mazzino,
  Seminara, and Tizzi}}]{celani}
\bibinfo{author}{\bibfnamefont{A.}~\bibnamefont{Celani}},
  \bibinfo{author}{\bibfnamefont{A.}~\bibnamefont{Mazzino}},
  \bibinfo{author}{\bibfnamefont{A.}~\bibnamefont{Seminara}}, \bibnamefont{and}
  \bibinfo{author}{\bibfnamefont{M.}~\bibnamefont{Tizzi}}, \bibinfo{journal}{J.
  Turbul.} \textbf{\bibinfo{volume}{8}} (\bibinfo{year}{2007}).

\bibitem[{\citenamefont{Celani et~al.}(2005)\citenamefont{Celani, Falkovich,
  Mazzino, and Seminara}}]{mazzino}
\bibinfo{author}{\bibfnamefont{A.}~\bibnamefont{Celani}},
  \bibinfo{author}{\bibfnamefont{G.}~\bibnamefont{Falkovich}},
  \bibinfo{author}{\bibfnamefont{A.}~\bibnamefont{Mazzino}}, \bibnamefont{and}
  \bibinfo{author}{\bibfnamefont{A.}~\bibnamefont{Seminara}},
  \bibinfo{journal}{Europhys. Lett.} \textbf{\bibinfo{volume}{70}},
  \bibinfo{pages}{775} (\bibinfo{year}{2005}).

\bibitem[{\citenamefont{Paoli and Shariff}(2009)}]{paoli}
\bibinfo{author}{\bibfnamefont{R.}~\bibnamefont{Paoli}} \bibnamefont{and}
  \bibinfo{author}{\bibfnamefont{K.}~\bibnamefont{Shariff}},
  \bibinfo{journal}{J. Atmos. Sci.} \textbf{\bibinfo{volume}{66}}
  (\bibinfo{year}{2009}).

\bibitem[{\citenamefont{Lanotte et~al.}(2009)\citenamefont{Lanotte, Seminara,
  and Toschi}}]{lanotte}
\bibinfo{author}{\bibfnamefont{A.~S.} \bibnamefont{Lanotte}},
  \bibinfo{author}{\bibfnamefont{A.}~\bibnamefont{Seminara}}, \bibnamefont{and}
  \bibinfo{author}{\bibfnamefont{F.}~\bibnamefont{Toschi}},
  \bibinfo{journal}{J. Atmos. Sci.} \textbf{\bibinfo{volume}{66}}
  (\bibinfo{year}{2009}).

\bibitem[{sup()}]{suppl}
\bibinfo{note}{See Supplemental Material [url], which includes
  Refs.\cite{siebert,schum2,vailyau,rogallo,pope88,smag}, for details on the
  physical models, further information on the numerical scheme, and additional
  results}.
  
\bibitem[{\citenamefont{Siebert et~al.}(2006)\citenamefont{Siebert, Lehmann,
  and Wendisch}}]{siebert}
\bibinfo{author}{\bibfnamefont{H.}~\bibnamefont{Siebert}},
  \bibinfo{author}{\bibfnamefont{K.}~\bibnamefont{Lehmann}}, \bibnamefont{and}
  \bibinfo{author}{\bibfnamefont{M.}~\bibnamefont{Wendisch}},
  \bibinfo{journal}{J. Atmos. Sci.} \textbf{\bibinfo{volume}{63}},
  \bibinfo{pages}{1451} (\bibinfo{year}{2006}).

\bibitem[{\citenamefont{Kumar et~al.}(2013)\citenamefont{Kumar, Schumacher, and
  Shaw}}]{schum2}
\bibinfo{author}{\bibfnamefont{B.}~\bibnamefont{Kumar}},
  \bibinfo{author}{\bibfnamefont{J.}~\bibnamefont{Schumacher}},
  \bibnamefont{and} \bibinfo{author}{\bibfnamefont{R.~A.} \bibnamefont{Shaw}},
  \bibinfo{journal}{Theor. Comp. Fluid Dyn.} \textbf{\bibinfo{volume}{27}},
  \bibinfo{pages}{361} (\bibinfo{year}{2013}).

\bibitem[{\citenamefont{Vaillancourt and Yau}(2000)}]{vailyau}
\bibinfo{author}{\bibfnamefont{P.~A.} \bibnamefont{Vaillancourt}}
  \bibnamefont{and} \bibinfo{author}{\bibfnamefont{M.}~\bibnamefont{Yau}},
  \bibinfo{journal}{B. Am. Metereolog. Soc.} \textbf{\bibinfo{volume}{81}},
  \bibinfo{pages}{285} (\bibinfo{year}{2000}).

\bibitem[{\citenamefont{Rogallo}(1981)}]{rogallo}
\bibinfo{author}{\bibfnamefont{R.}~\bibnamefont{Rogallo}},
  \bibinfo{journal}{NASA STI/Recon Technical Report N}
  \textbf{\bibinfo{volume}{81}}, \bibinfo{pages}{31508} (\bibinfo{year}{1981}).

\bibitem[{\citenamefont{Eswaran and Pope}(1988)}]{pope88}
\bibinfo{author}{\bibfnamefont{V.}~\bibnamefont{Eswaran}} \bibnamefont{and}
  \bibinfo{author}{\bibfnamefont{S.}~\bibnamefont{Pope}},
  \bibinfo{journal}{Comput. Fluids} \textbf{\bibinfo{volume}{16}},
  \bibinfo{pages}{257} (\bibinfo{year}{1988}).

\bibitem[{\citenamefont{Smagorinsky}(1963)}]{smag}
\bibinfo{author}{\bibfnamefont{J.}~\bibnamefont{Smagorinsky}},
  \bibinfo{journal}{Mon. Weather Rev.} \textbf{\bibinfo{volume}{91}},
  \bibinfo{pages}{99} (\bibinfo{year}{1963}).

\bibitem[{\citenamefont{Twomey}(1959)}]{twom}
\bibinfo{author}{\bibfnamefont{S.}~\bibnamefont{Twomey}},
  \bibinfo{journal}{Pure Appl. Geophys.} \textbf{\bibinfo{volume}{43}},
  \bibinfo{pages}{243} (\bibinfo{year}{1959}).

\bibitem[{\citenamefont{Pruppacher and Klett}(1997)}]{prup}
\bibinfo{author}{\bibfnamefont{H.}~\bibnamefont{Pruppacher}} \bibnamefont{and}
  \bibinfo{author}{\bibfnamefont{J.}~\bibnamefont{Klett}},
  \emph{\bibinfo{title}{Microphysics of clouds and precipitation}}
  (\bibinfo{year}{1997}).

\bibitem[{\citenamefont{Pope}(2000)}]{pope}
\bibinfo{author}{\bibfnamefont{S.~B.} \bibnamefont{Pope}},
  \emph{\bibinfo{title}{Turbulent flows}} (\bibinfo{publisher}{Cambridge
  university press}, \bibinfo{year}{2000}).

\bibitem[{\citenamefont{Kumar et~al.}(2014)\citenamefont{Kumar, Schumacher, and
  Shaw}}]{schum}
\bibinfo{author}{\bibfnamefont{B.}~\bibnamefont{Kumar}},
  \bibinfo{author}{\bibfnamefont{J.}~\bibnamefont{Schumacher}},
  \bibnamefont{and} \bibinfo{author}{\bibfnamefont{R.~A.} \bibnamefont{Shaw}},
  \bibinfo{journal}{J. Atmos. Sci.} \textbf{\bibinfo{volume}{71}},
  \bibinfo{pages}{2564Ð2580} (\bibinfo{year}{2014}).

\bibitem[{\citenamefont{Zhan et~al.}(2014)\citenamefont{Zhan, Sardina, Lushi,
  and Brandt}}]{cj}
\bibinfo{author}{\bibfnamefont{C.}~\bibnamefont{Zhan}},
  \bibinfo{author}{\bibfnamefont{G.}~\bibnamefont{Sardina}},
  \bibinfo{author}{\bibfnamefont{E.}~\bibnamefont{Lushi}}, \bibnamefont{and}
  \bibinfo{author}{\bibfnamefont{L.}~\bibnamefont{Brandt}},
  \bibinfo{journal}{J. Fluid Mech.} \textbf{\bibinfo{volume}{739}},
  \bibinfo{pages}{22} (\bibinfo{year}{2014}).

\bibitem[{\citenamefont{Ditas et~al.}(2012)\citenamefont{Ditas, Shaw, Siebert,
  Simmel, Wehner, and Wiedensohler}}]{ditas}
\bibinfo{author}{\bibfnamefont{F.}~\bibnamefont{Ditas}},
  \bibinfo{author}{\bibfnamefont{R.}~\bibnamefont{Shaw}},
  \bibinfo{author}{\bibfnamefont{H.}~\bibnamefont{Siebert}},
  \bibinfo{author}{\bibfnamefont{M.}~\bibnamefont{Simmel}},
  \bibinfo{author}{\bibfnamefont{B.}~\bibnamefont{Wehner}}, \bibnamefont{and}
  \bibinfo{author}{\bibfnamefont{A.}~\bibnamefont{Wiedensohler}},
  \bibinfo{journal}{Atmos. Chem. Phys.} \textbf{\bibinfo{volume}{12}},
  \bibinfo{pages}{2459} (\bibinfo{year}{2012}).



\end{thebibliography}
\end{document}


\title{Supplemental material for ``Continuous growth of droplet size variance due to condensation in turbulent cloud''}

\author{Gaetano Sardina}\email[]{gaetano.sardina@misu.su.se}
\affiliation{Department of Meteorology and SeRC, Stockholm University, Stockholm, Sweden}
\author{Francesco Picano}
\affiliation{Linn{\'e} FLOW Centre and SeRC, KTH Mechanics, Stockholm, Sweden}
\affiliation{Department of Industrial Engineering, University of Padova, Padova, Italy}
\author{Luca Brandt}
\affiliation{Linn{\'e} FLOW Centre and SeRC, KTH Mechanics, Stockholm, Sweden}
\author{Rodrigo Caballero}
\affiliation{Department of Meteorology and SeRC, Stockholm University, Stockholm, Sweden}

\date{\today}



\maketitle

\section{SECTION A. EULERIAN-LAGRANGIAN PHYSICAL CLOUD MODEL}

\subsection{Derivation of the supersaturation equation}

The equation governing the evolution of the supersaturation can be derived from first principles, vapor mass and energy conservation, and thermodynamics relations. Since the supersaturation  is a thermodynamic  variable (as temperature, internal energy and entropy) one can write a conservation equation. The original  derivation by Twomey that can be found in  textbooks of atmospheric physics is obtained neglecting the effects of viscosity \cite{twom,prup}, which we include below in the most general case.

The supersaturation $s$ is defined as $s=q_v/q_{vs}-1$ where $q_v$ is the vapor mixing ratio and $q_{vs}$ its saturated value. The supersaturation rate of change is obtained by differentiating its definition:
\begin{equation}
\frac{Ds}{Dt}=\frac{1}{q_{vs}}\frac{Dq_v}{Dt}-\frac{q_v}{q^2_{vs}}\frac{Dq_{vs}}{Dt}
\label{first}
\end{equation}
The vapor mixing ratio follows the conservation equation:
\begin{equation}
\frac{Dq_v}{Dt}=D_{v}\nabla^2 q_v-C_d
\end{equation}
where $D_v$ is the thermal diffusivity of water vapor in air and $C_d$ is the condensation rate.
An evolution  equation for $q_{vs}$ is obtained from the Clasius-Clapeyron equation linking supersaturation pressure $e_s$ with temperature $\theta$:
\begin{align}
\frac{Dq_{vs}}{Dt}=\frac{D}{Dt} \left (  \frac{\epsilon e_{s}}{p} \right )=\frac{\epsilon}{p}\frac{D e_s}{Dt}-\frac{\epsilon e_s}{p^2}\frac{D p}{Dt}= \nonumber \\
=\frac{\epsilon}{p}\frac{d e_s}{d \theta}\frac{D \theta}{D t}-\frac{\epsilon e_s}{p^2}\frac{D p}{Dt}=\frac{\epsilon L e_s}{p R_v \theta^2}\frac{D \theta}{D t}-\frac{\epsilon e_s}{p^2}\frac{D p}{Dt}
\label{sat}
\end{align}
where $\epsilon$ is the ratio between water and dry air molecular weights, $p$ is the thermodynamic pressure, $R_v$ is the gas costant for water vapor and $L$ is the latent heat of evaporation.
The temperature $\theta$ obeys energy conservation:
\begin{equation}
\frac{D \theta}{D t}= \kappa_\theta \nabla^2 \theta -\frac{g}{c_p} w+\frac{L}{c_p}C_d 
\label{tem}
\end{equation}
where $k_\theta$ is the thermal diffusivity, $g$ the gravity acceleration, $c_p$ is the specific heat at costant pressure and $w$ the vertical fluid velocity component.
The pressure $p$ can be obtained from hydrostatic equilibrium:
\begin{equation}
\frac{D p}{D t}=-\frac{gp}{R_a \theta} w 
\label{pre}
\end{equation}
where $R_a$ is the gas costant for dry air.
Substituting Eqs. (\ref{pre}) and (\ref{tem}) in Eq. (\ref{sat}) results in:
\begin{equation}
\frac{Dq_{vs}}{Dt}=\left (\frac{\epsilon  e_s g}{p R_a \theta}-\frac{\epsilon  e_s g L}{p R_v \theta^2 c_p} \right ) w+\frac{\epsilon  e_s k_\theta L}{p R_v \theta^2}\nabla^2 \theta+\frac{\epsilon  e_s  L^2}{p R_v \theta^2 c_p}C_d
\end{equation}
and consequently Eq. (\ref{first}) becomes:
\begin{align}
&\frac{D s}{Dt}=\frac{1}{q_{vs}}D_v\nabla^2 q_v-\frac{C_d}{q_{vs}}+\frac{s+1}{q_{vs}}\left (\frac{\epsilon  e_s g L}{p R_v \theta^2 c_p}- \frac{\epsilon  e_s g}{p R_a \theta}  \right) w +\nonumber \\
&-\frac{s+1}{q_{vs}}\frac{\epsilon  e_s k_\theta L}{p R_v \theta^2}\nabla^2 \theta-\frac{s+1}{q_{vs}}\frac{\epsilon  e_s  L^2}{p R_v \theta^2 c_p}C_d .
\end{align}
Assuming $s\ll1$ and substituting the expression of $q_{vs}$, the previous equation reads:
\begin{align}
&\frac{D s}{Dt}=\left (\frac{g L}{R_v \theta^2 c_p}- \frac{g}{R_a \theta}  \right) w-\left (\frac{p}{\epsilon e_s}+\frac{L^2}{R_v c_p \theta^2} \right )C_d\nonumber \\
&+\frac{p D_v}{\epsilon e_s}\nabla^2 q_v-\frac{k_\theta L}{R_v \theta^2}\nabla^2 \theta ,
\label{sup1}
\end{align}
where we defined
\begin{align}
A_1(\theta)&=\frac{gL}{R_vc_p\theta^2}-\frac{g}{R_a\theta}\\
A_2(\theta)&=\frac{R_a\theta}{\epsilon e_s(\theta)}+\frac{\epsilon L^2}{p\theta c_p}\\
A_3(\theta)&=\left (\frac{\rho_w R_v \theta}{D_v e_s(\theta)}+\frac{\rho_w L^2}{k_\theta R_v \theta^2} \right )^{-1}
\end{align}
and  
\begin{equation}
C_d=\frac{4\pi \rho_w A_3}{\rho V} s \sum R_i 
\end{equation}
The supersaturation relaxation time is 
\begin{equation}
\tau_s(\theta)=\left (4\pi \rho_w A_2(\theta) A_3(\theta) \sum R_i/V \right )^{-1}
\end{equation}
where $\rho_w$ is the water density and $R_i$ are the radii of the droplets in the volume $V$. 
 Eq. (\ref{sup1}) can be exactly re-written as
\begin{equation}
\frac{D s}{Dt}=A_1 w-\frac{s}{\tau_s}+\frac{p D_v}{\epsilon e_s}\nabla^2 q_v-\frac{k_\theta L}{R_v \theta^2}\nabla^2 \theta.
\label{sup2}
\end{equation}
 
The only assumption in the model is introduced now, assuming the diffusive terms proportional to the laplacian of the supersaturation
\begin{equation}
k\nabla^2s=\frac{p D_v}{\epsilon e_s}\nabla^2 q_v-\frac{k_\theta L}{R_v \theta^2}\nabla^2 \theta
\end{equation}
obtaining the final equation for the supersaturation field
\begin{equation}
\frac{D s}{Dt}=A_1 w-\frac{s}{\tau_s}+k\nabla^2 s.
\label{sup3}
\end{equation}
Given the high Reynolds numbers inside clouds it appears reasonable to assume that 
diffusion and viscous effects are much smaller than the convective transport.
Indeed, we will show in Section B that the diffusive term does not affect the supersaturation for large cloud dimensions. 
The term $A_1 w$ is a source/sink term of supersaturation resulting from the variation in temperature and pressure with height.

\subsection{Model equations and parameters}

The complete Eulerian set of equations for the turbulent velocity field $\bf u$, the supersaturation $s$, the temperature fluctuations $\theta$, the vapor mixing ratio $q_v$ and the buoyancy $B$  therefore reads
\begin{align}
\nabla \cdot {\bf u}&=0,\label{massa}\\
\frac{\partial {\bf u}}{\partial t}+{\bf u}\cdot\nabla {\bf u}&=-\frac{\nabla p'}{\rho}+\nu \nabla^2 {\bf u}+B{\bf e_z}+ {\bf f}, \label{fluid_u}\\
\frac{\partial s}{\partial t}+{\bf u}\cdot\nabla s &= \kappa \nabla^2 s + A_1(\theta) w -\frac{s}{\tau_s(\theta)}, \label{fluid_s}\\
\frac{\partial \theta}{\partial t}+{\bf u}\cdot\nabla \theta &= \kappa_\theta \nabla^2 \theta -\frac{g}{c_p} w +\frac{L}{c_p}C_d, \label{fluid_T}\\
q_v&=q_{vs}(\theta) (s+1),\label{fluid_q}\\
B&=g\left [\frac{\theta-\theta_0}{\theta_0}+\epsilon (q_v-q_{v0})-q_w \right ]. \label{fluid_B}
\end{align}
The velocity field $\bf u$ obeys the Navier-Stokes-Boussinesq equations, (\ref{fluid_u}),
where $p'$ is the pressure deviation from its hydrostatic value, $\rho$ is the 
density of air, ${\bf f}$ an external forcing acting just at the larger turbulent scales, 
$\nu$ the kinematic viscosity.
The external forcing ${\bf f}$ maintains a turbulent field in the steady state regime. 

In the most general cases turbulent clouds are a transient phenomenon. The transient effects are relevant when observing the cloud for its whole life-time, especially during the cloud generation and dissipation phase. 
The total life-time of a cloud is of the  order of hours/days;
however, if  we study droplet condensation for time intervals of the order of minutes,
we can consider the system at quasi statistical steady-state, as shown by experimental observations in real clouds as shallow cumulus clouds \cite{siebert} and marine stratocumulus clouds \cite{ditas}. Most of the DNS of clouds  available in literature assume steady state turbulence \cite{vail,paoli,lanotte,schum,schum2}.
The temperature follows the advection-diffusion equation (\ref{fluid_T}).
Equation (\ref{fluid_q}) links the vapor mixing ratio $q_v$, with its saturated value $q_{vs}=q_{vs}(\theta,s)$ function of the temperature and the supersaturation field.
The momentum equation is forced by the buoyancy field $B$, expressed by equation (\ref{fluid_B}), where $\theta_0$ is the reference temperature, $q_{v0}$ is the  vapor mixing ratio at $\theta_0$ and $s=0$, $q_w$ is the water mixing ratio and represents the droplet drag on the flow. In adiabatic cloud cores, the buoyancy effects are assumed to be negligible \cite{vailyau}.

The Eulerian fields are coupled with the Lagrangian equation for the droplet evolution:
\begin{align}
\frac{d {\bf v_d}}{dt}&=\frac{{\bf u}({\bf x_d},t)-{{\bf v_d}}}{\tau_d}-g{\bf e_z},\label{vel_d}\\
\frac{d {\bf x_d}}{dt}&={\bf v_d},\label{space_d}\\
\frac{d { R_i^2}}{dt}&=2 A_3(\theta) {s({\bf x_d},t)} \label{rad_d}
\end{align}
with ${\bf x_d}$ and ${\bf v_d}$ the droplet position and velocity, ${\bf u}({\bf
  x_d},t)$ and $s({\bf x_d},t)$ the fluid velocity and supersaturation field at droplet position, $\tau_d=2\rho_w R_i^2/(9 \rho\nu)$ the droplet relaxation time.
The droplets are assumed to be spherical, with radius smaller than the Kolmogorov scale in order to be considered point-particles. The only forces acting on the droplets are the Stokes drag and the gravitational force.
\subsection{Computational methodology}

The  numerical  data  set  described in the manuscript has  been  obtained  from  Direct Numerical Simulations (DNS)  using  a  classical  pseudo-spectral  method  for the Eulerian equations coupled  with  a  Lagrangian  solver  for  the  droplets.
For  the  fluid  phase,  the  Navier--Stokes--Boussinesq  equations  have  been  integrated  in  a  three-
periodic domain using a Fourier spectral method with the nonlinear terms  de-aliased  by  the  $3/2$  rule.  The  solution  is  advanced  in  time  using  a  third-order low-storage Runge--Kutta method; specifically, the nonlinear terms are computed using an Adam--Bashforth-like approximation while the diffusive terms are analytically integrated  \cite{rogallo}.  The  random  forcing  $\bf f$ is  applied  isotropically  to  the  first  shell of  wave  vectors,  with  fixed  amplitude, constant  in  time  and  uniformly  distributed
in  phase  and  directions \cite{pope88}.  
For the dispersed phase, we use the point-particle approximation as typical droplet sizes are smaller than the flow Kolmogorov length. The same Runge--Kutta
temporal  integration  used  for  the  carrier  phase  is  adopted  for  the  Lagrangian equations.  
A  trilinear  interpolation  scheme  is  used  to  compute  the  flow  velocity, supersaturation and temperature at the  particle  position. The correspondent trilinear extrapolation scheme is adopted to transfer the information from the droplet location (supersaturation, radius, temperature) into the Eulerian grid.
The code is fully MPI parallelized to be efficiently run on supercomputing infrastructures with a performance linearly scaling up to $10^4$ cores.

\section{SECTION B. NUMERICAL MODELS}

\subsection{Constant coefficient model}
\begin{figure}
\centering
\includegraphics[width =.49\textwidth]{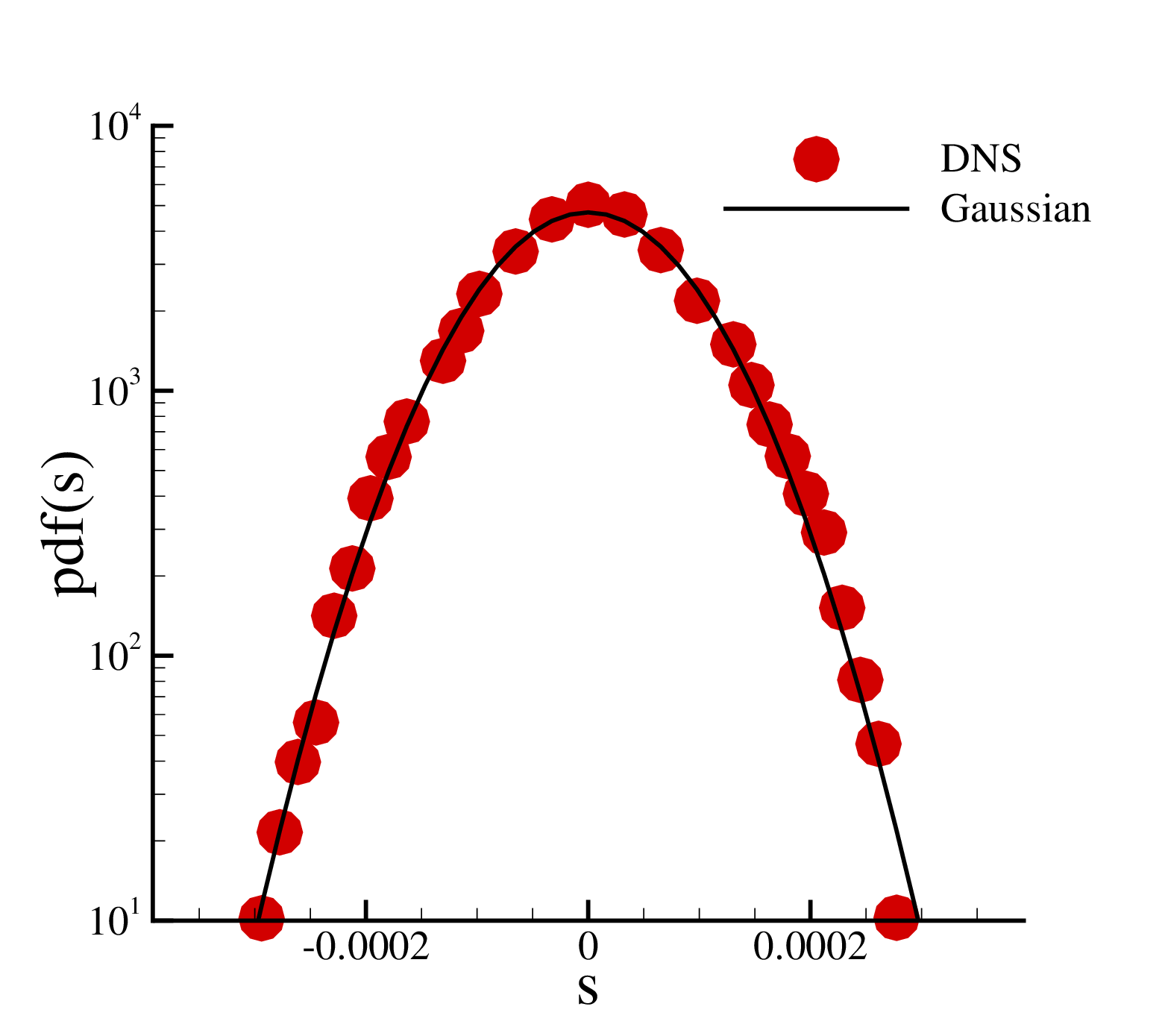}
\caption{Probability distribution function of the supersaturation fluctuations extracted by the DNS simulation (symbols) and comparison with its Gaussian fit (line).}
\label{fig_1}
\end{figure}

The system of differential and algebraic equations (1-14) can be simplified assuming that 
the coefficients $A_1$, $A_2$, $A_3$ are only function of the reference temperature $\theta_0$, i.e.\ the temperature fluctuations can be neglected for their calculation \cite{lanotte,schum2}, and  the buoyancy term can be neglected as well \cite{vailyau}.
To check and quantify the order of magnitude of the temperature fluctuations, we performed a preliminary DNS simulation solving the complete system (1-14).
The simulation matches the parameters described in  case DNS $D1$ of the manuscript. In particular, we simulate a cloud of length $L_{box}=1.5$ m, velocity fluctuations $v_{rms}=0.1$ m/s, Taylor Reynolds number $Re_\lambda=390$ with a  resolution of $1024^3$ grid points and $4.4\times10^8$ Lagrangian droplets. From the numerical results we obtain a negligible value of the temperature root mean square $\theta_{rms}=5\times 10^{-3}$ K. Since, the supersaturation depends on a non-trivial way on the turbulent field via the quantities $A_2$, $A_3$, $\tau_s$ and the Clasius-Clapeyron equation, it is relevant to examine the behavior of its probability distribution function (pdf).
Figure \ref{fig_1} shows the pdf of the supersaturations fluctuations for the full set of equations (DNS): the behavior is Gaussian and equal to the two results from the constant coefficient model reported in the manuscript. 

\subsection{Direct Numerical Simulation (DNS) of the reduced model}
As seen in the previous paragraph, we can assume that the temperature field is constant in time and space and equals to its reference value $\theta_0$ for the calculation of the constants $A_1$, $A_2$ and $A_3$ while the fluctuations still affect the evaporation-condensation via the supersaturation equation. This has two consequences: the physics of the system is simpler to investigate and the numerical solver is faster since the equations are more easy to manage. Under the simplifying hypothesis the system (1-14) reduces to:

\begin{align}
\nabla \cdot {\bf u}&=0,\label{massa1}\\
\frac{\partial {\bf u}}{\partial t}+{\bf u}\cdot\nabla {\bf u}&=-\frac{\nabla p'}{\rho}+\nu \nabla^2 {\bf u}+ {\bf f}, \label{fluid_u1}\\
\frac{\partial s}{\partial t}+{\bf u}\cdot\nabla s &= \kappa \nabla^2 s + A_1 w -\frac{s}{\tau_s}, \label{fluid_s1}\\
\frac{d {\bf v_d}}{dt}&=\frac{{\bf u}({\bf x_d},t)-{{\bf v_d}}}{\tau_d}-g{\bf e_z},\label{vel_d1}\\
\frac{d {\bf x_d}}{dt}&={\bf v_d},\label{space_d1}\\
\frac{d { R_i^2}}{dt}&=2 A_3 {s({\bf x_d},t)} \label{rad_d1}
\end{align}

where $A_1$, $A_2$ and $A_3$ are constant that depends on the reference temperature $\theta_0$.
The system (15-20) can be numerically solved  without any further hypothesis, simplifications or models via Direct Numerical Simulations. In the manuscript we performed 8 different DNS cases described, see Table I, the largest corresponding to a cloud of the order of $L_{box}=3$ m (DNS E1) with a resolution of $10^{10}$ degrees of freedom including the Eulerian grid points ($2048^3$) and Lagrangian droplets ($3\times 10^9$). 

\subsection{Large Eddy Simulation (LES)}

Since the largest DNS simulation allows us to describe core clouds of the order of 3 meters, we need to employ Large Eddy Simulations (LES) to study clouds with size of about 100 meters. 
The LES equations are obtained from the DNS system by applying a low-pass filter to the Eulerian fields that remove the smallest scales. In this way, just the large scales are evolved while the effects of the small scales on the large ones is modeled. The generic Eulerian field $f$ can be decomposed in a
filtered or resolved component $\bar f$ and a residual or subgrid-scale component $f_{sgs}$.
Applying the filter operator to the system (15-20), we obtain:
\begin{align}
\nabla \cdot {\bar{\bf u}}&=0,\label{massa1}\\
\frac{\partial {\bar{\bf u}}}{\partial t}+{\bar{\bf u}}\cdot\nabla {\bar{\bf u}}&=-\frac{\nabla \bar{p}}{\rho}+\nu \nabla^2 {\bar{\bf u}}+ {\bar{\bf f}}+\nabla \cdot {\bar{\bf \tau}_{sgs}}, \label{fluid_u1}\\
\frac{\partial \bar s}{\partial t}+{\bar{\bf u}}\cdot\nabla \bar s &= \kappa \nabla^2 \bar s + A_1 \bar w -\frac{\bar s}{\tau_s}+\nabla \cdot {\bar{\bf J}_{sgs}}, \label{fluid_s1}\\
\frac{d {\bf v_d}}{dt}&=\frac{{\bar{\bf u}}({\bf x_d},t)+{\bf u_{sgs}}({\bf x_d},t)-{{\bf v_d}}}{\tau_d}-g{\bf e_z},\label{vel_d1}\\
\frac{d {\bf x_d}}{dt}&={\bf v_d},\label{space_d1}\\
\frac{d { R_i^2}}{dt}&=2 A_3 \left ({\bar s({\bf x_d},t)}+ s_{sgs}({\bf x_d},t)\right ) \label{rad_d1}
\end{align}

where ${\bar{\bf \tau}_{sgs}}$, ${\bar{\bf J}_{sgs}}$, ${\bf u_{sgs}}$ and $ s_{sgs}$ are respectively
the subgrid stress tensor, the subgrid supersaturation flux, the subgrid velocity and the subgrid supersaturation.
The subgrid terms should be described using suitable closure models, which introduce an unavoidable approximation of the physics with respect to the DNS.
Here, the subgrid stress tensor ${\bar{\bf \tau}_{sgs}}$ and the subgrid supersaturation flux ${\bar{\bf J}_{sgs}}$ are modelled with to the classic Smagorinski closure \cite{smag}.
The subgrid terms appearing in the Lagrangian droplet equation ${\bf u_{sgs}}$ and $ s_{sgs}$ are set  to zero so to consider our LES results as a lower limit for the droplet size distribution in the cloud.
Infact, considering the temporal evolution of  the droplet square radius fluctuations:
\begin{equation}
\frac{d { \langle (R_i^{2'})^2\rangle}}{dt}=\frac{d \sigma_{R^2}^2}{dt}=4 A_3 \left (\langle {\bar s}' R^{2'}\rangle +\langle {s'_{sgs}} R^{2'}\rangle \right ) \label{exact_1}
\end{equation}
we are implicitly assuming that the subgrid correlations $\langle {s'_{sgs}} R^{2'}\rangle$ are much smaller than the resolved scale correlations $\langle {\bar s}' R^{2'}\rangle$. This is justified by the DNS results represented in the inset of figure 1 of the main manuscript where the correlation $\langle {s}' R^{2'}\rangle$ is displayed. The figure shows that the large DNS gives values of this correlation two order of magnitude larger than that from the smaller simulations.

\begin{figure}
\centering
\includegraphics[width =.49\textwidth]{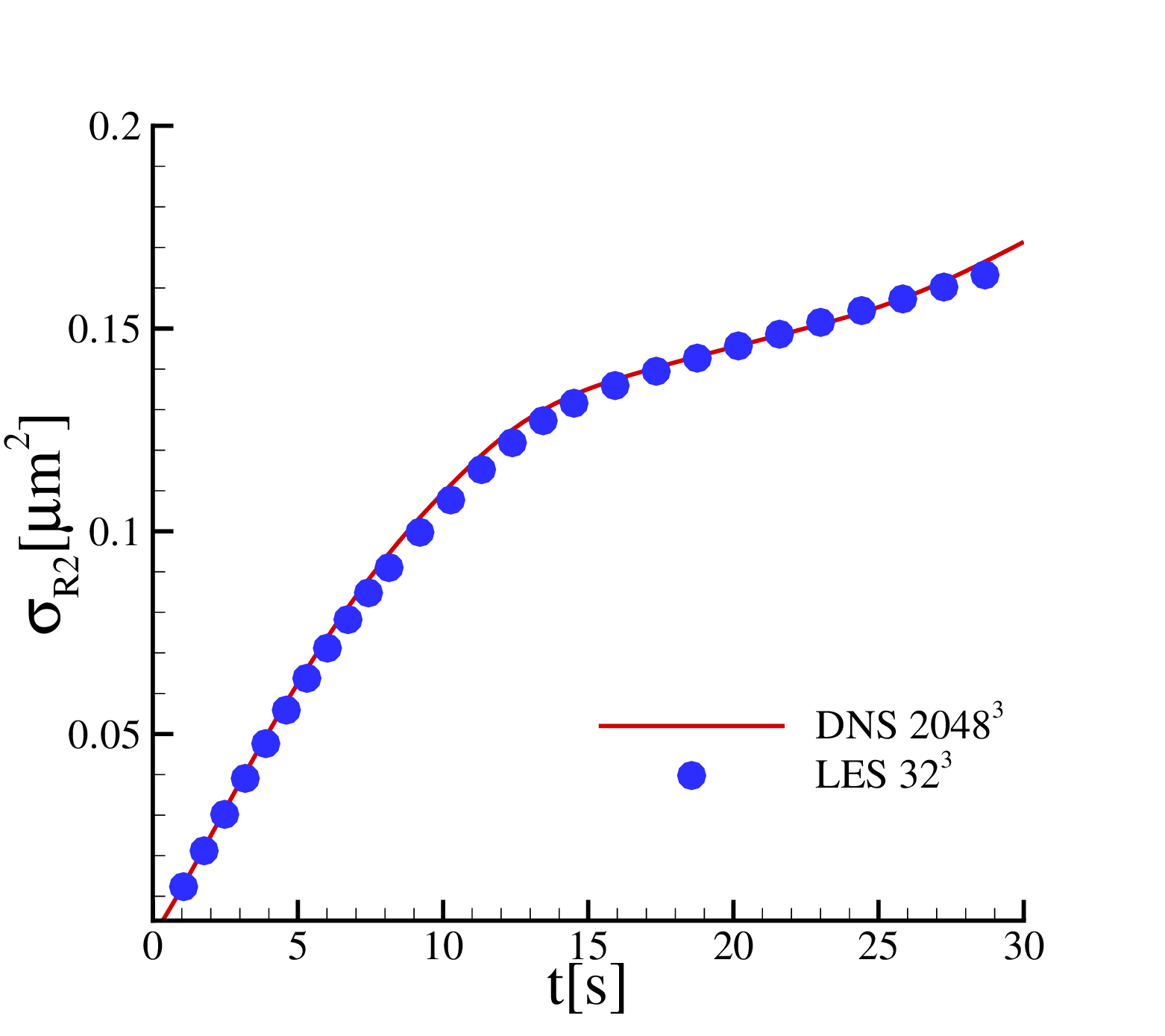}\\
\includegraphics[width =.49\textwidth]{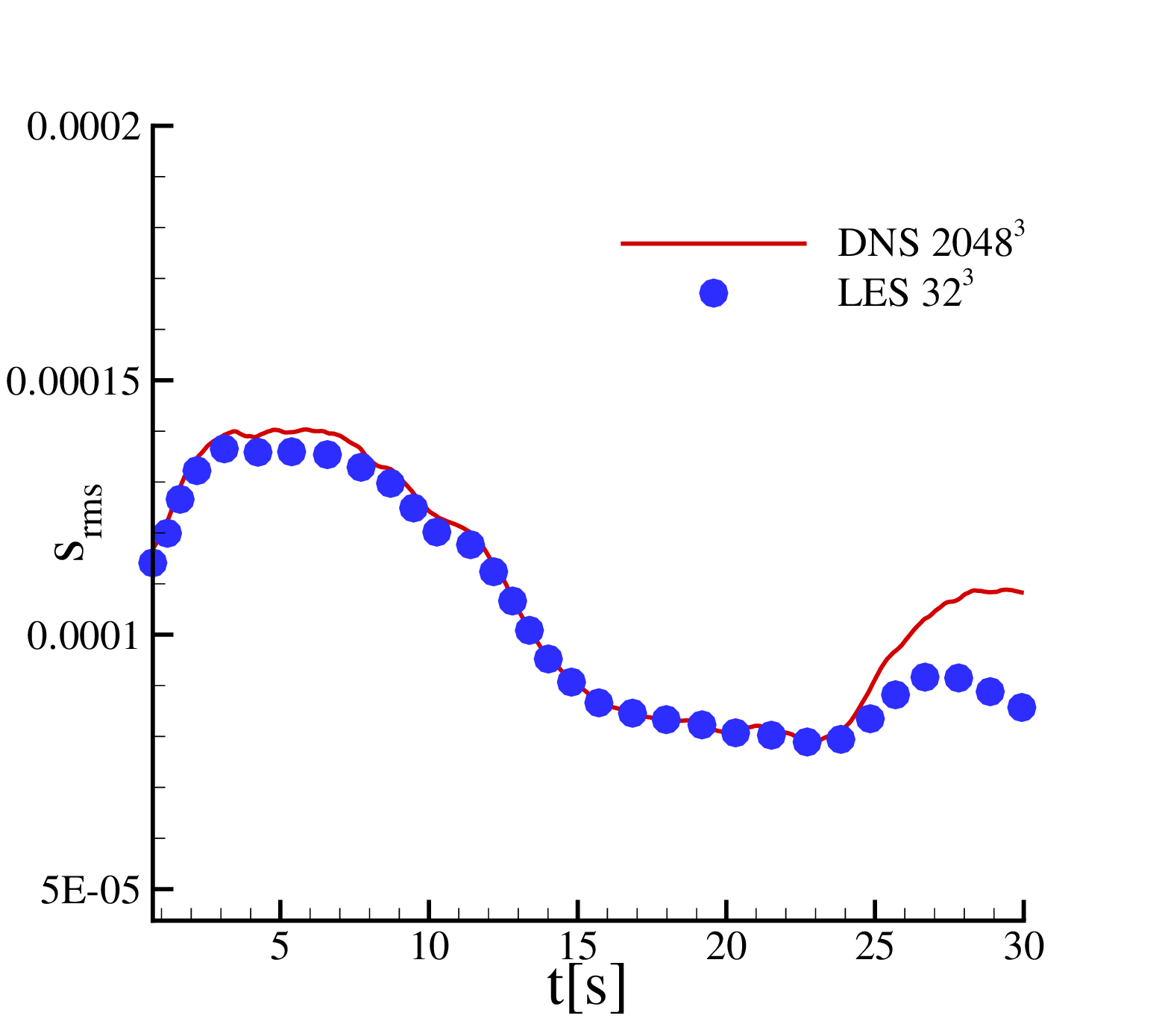}
\caption{Top panel: comparison between the root mean square of the square droplet radius fluctuations $\sigma_{R^2}$ versus time for the LES  (symbols) and the DNS simulations. Bottom panel: comparison between the root mean square of the supersaturation fluctuations versus time for the LES (symbols) and the DNS simulations.}
\label{fig_2}
\end{figure}

Another approximation introduced in the LES is related to the number of droplets tracked in the simulation.
Theoretically, one should evolve order $10^{14}$ droplets which is unfeasible. We therefore use 
the method of droplet renormalization described in
\cite{lanotte}.
We validated the LES by comparing the results with our finest DNS simulation, DNS E1.
Different LES simulations have been performed with different resolutions down to $32^3$ grid points. 
The root mean square of the square droplet radius fluctuations (top panel) and
of the supersaturation fluctuations (bottom panel) are shown in figure \ref{fig_2} for
the DNS and LES. The LES slightly underestimates the observable $\sigma_{R^2}^2$  and $s_{rms}$ as expected from equation (\ref{exact_1}). 
The agreement between LES and DNS is excellent considering that we used $10^5$ degrees of freedom against $10^{10}$ for the DNS. This implies that just solving the larger scales is enough to provide a good prediction of the droplet size spectra variance.
In conclusion, the LES simulation is able to give a lower limit of the droplet spectra broadening at a moderate computational cost.

\subsection{Stochastic model (SM)}
The other  approach we used in our work is the stochastic model described by the equations (6-7) in the main manuscript. 
We show in the main text that the stochastic model tends to overestimate the droplet-related quantities  and that the differences become smaller by increasing the ratio between the large and the small scales in the numerical simulations.
A systematic study of the difference between the stochastic model and the DNS
is presented  in this section.
\begin{figure}
\centering
\includegraphics[width =.49\textwidth]{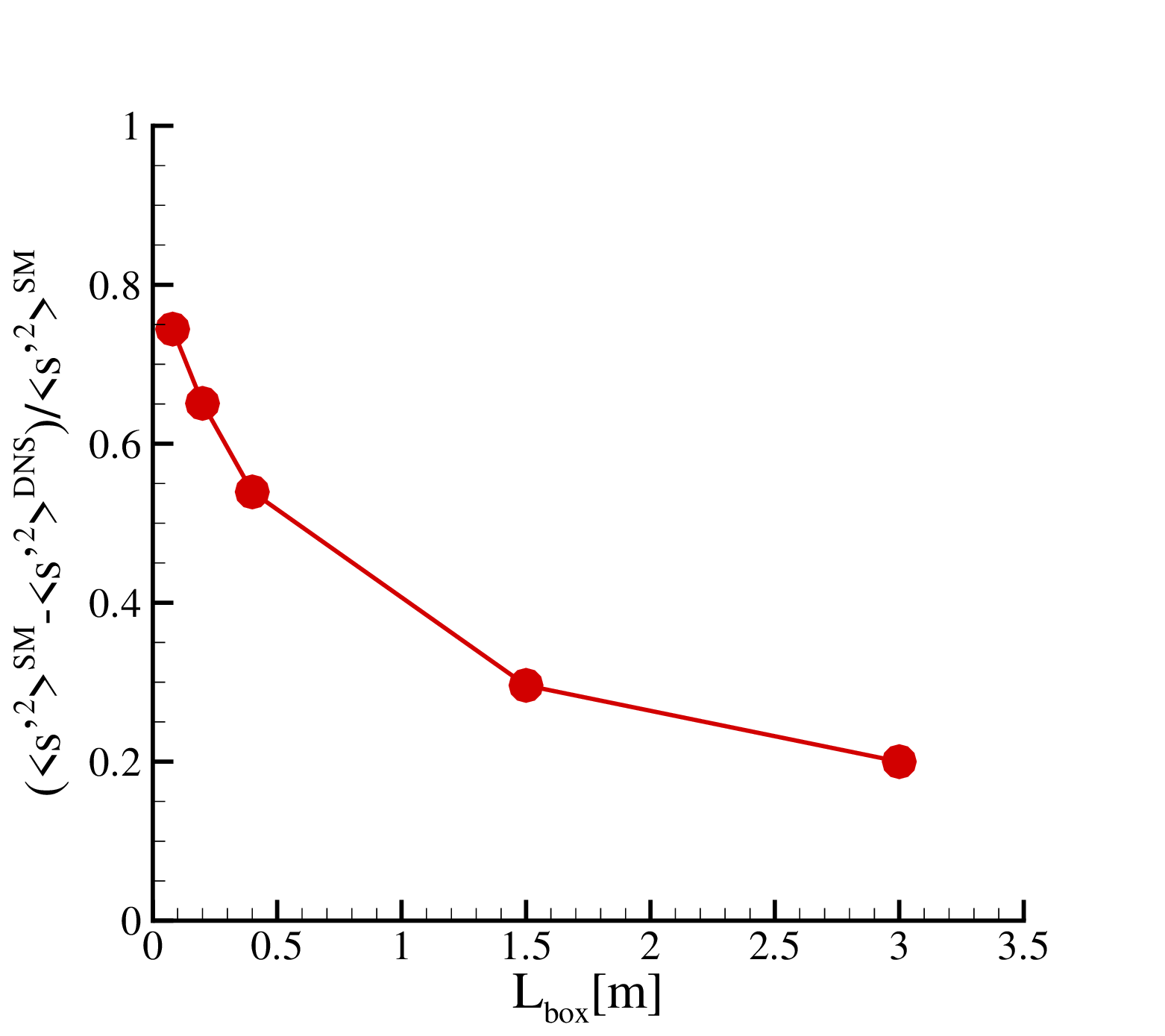}\\
\includegraphics[width =.49\textwidth]{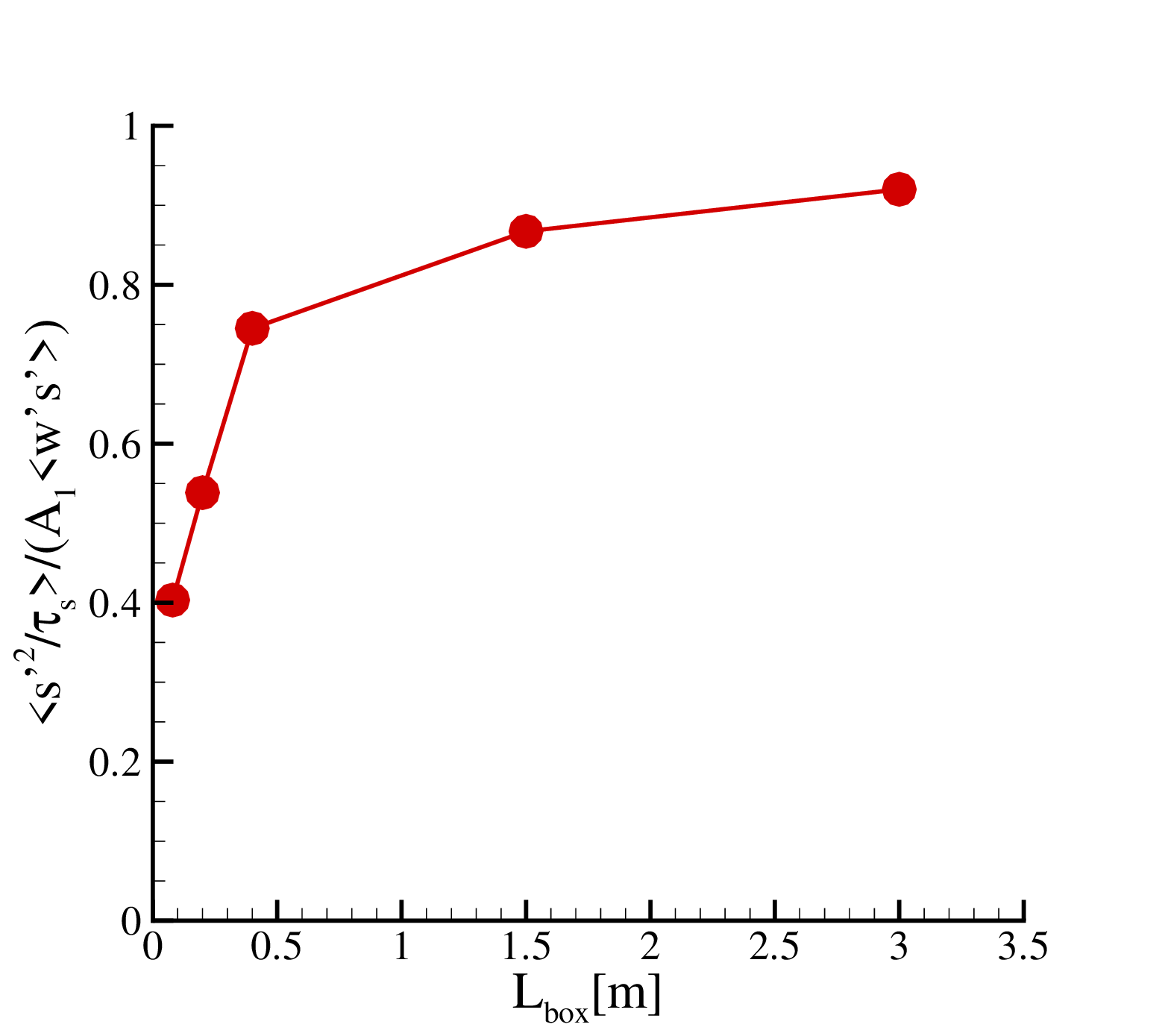}
\caption{Top panel: Relative error pertaining   the quasi steady-state supersaturation fluctuation variance in the stochastic model (SM) and  DNS. Bottom panel: ratio between the droplet sink term and the source term in the supersaturation variance equation (28). }
\label{fig_3}
\end{figure}

The relative  error pertaining the quasi steady-state value of the supersaturation fluctuation variance is reported in the top panel of figure \ref{fig_3} 
versus the cloud size domain (the size of the computational domain and the large-to-small scale ratio).
A large error is evident for clouds of smaller size and  the differences are monotonically decreasing when increasing the large-to-small scale ratio (i.e. the Reynolds number). 
For clouds that are of the order of $100$ meters, our prediction has a margin of error lower than  $20\;\%$, an acceptable value
for statistical observables that DNS cannot calculate with the current computational resources.

The error in the stochastic model can be explained by viscous effects in the supersaturation fluctuations.
Multiplying equation (17) with the supersaturation fluctuations and averaging, we obtain an equation for the
supersaturation variance:
\begin{equation}
\frac{\partial \langle s'^2/2 \rangle}{\partial t}+\nabla\cdot \langle {\bf u'}s'^2/2\rangle = -\kappa \langle \left (\nabla s'\right )^2 \rangle+ A_1 \langle w's' \rangle -\langle \frac{s'^2}{\tau_s} \rangle
\end{equation}
where the left hand side is zero at statistically steady state, and a balance between three terms holds:
viscous dissipation $\kappa \langle \left (\nabla s'\right )^2 \rangle$, production $A_1 \langle w's' \rangle$ and droplet sink  $\langle {s'^2}/{\tau_s} \rangle$. In the case of the classic inviscid Twomey equation (employed also in the stochastic model) the balance is only between production and the droplet sink terms, and viscous effects are neglected. 
To quantify the differences between the inviscid and viscid Twomey equation, we display the ratio between the droplet sink and the production term as extracted from the DNS
in the bottom panel of figure \ref{fig_3}, where a value of this ratio equal to 1 corresponds to an inviscid behavior as assumed in the Towmey model.
For smaller clouds, the ratio is about $0.5$ so that just half of the supersaturation fluctuations induce the droplet growth while the remaining is diffused. This explain why the stochastic model, neglecting diffusion, is not able to predict the DNS results. By increasing the 
Reynolds number 
and the size of the cloud, the ratio $\langle {s'^2}/{\tau_s} \rangle/ A_1 \langle w's' \rangle $
tends to 1, meaning that the viscous effects become more and more negligible (they vanish in the limit of infinite Reynolds number). 
Therefore, the  stochastic model effectively predicts an upper limit for the droplet spectral broadening at a zero computational cost.
In conclusion, the DNS results will always lie between the LES estimation (lower limit) and the stochastic model prediction (upper limit).